\documentstyle[12pt]{article}
\input{epsfig.sty}
\newcommand{\beq}{\begin{eqnarray}}
\newcommand{\eeq}{\end{eqnarray}}
\begin{document}
\title{$\Psi$ and $\Upsilon$ production via $\sqrt{s_{pp}}$=8 TeV p-Pb 
collisions}
\author{Leonard S. Kisslinger\\
Department of Physics, Carnegie Mellon University, Pittsburgh, PA 15213}

\date{}
\maketitle
\begin{abstract}
   We estimate differential rapidity cross sections for $\Psi$ and $\Upsilon$ 
production via p-Pb collisions at 8 TeV. We use the mixed heavy quark hybrid
theory in which the $J/\Psi(1S),\Upsilon(1S),\Upsilon(2S)$ are standard
mesons while the $\Psi(2S)$ and $\Upsilon(3S)$ are mixed hybrids, approximately
50\% standard $|q \bar{q}>$ states and 50\% hybrid $|q \bar{q} g>$ states. 
This is an extension of previous work on heavy-quark state production via
A-A collisions at RHIC.
\end{abstract}
\noindent
PACS Indices:12.38.Aw,13.60.Le,14.40.Lb,14.40Nd
\vspace{1mm}

\section{Introduction}

  Our previous work\cite{klm14} on the production of  $\Psi$ and $\Upsilon$ 
state production via Cu-Cu and Au-Au collisions, with E=$\sqrt{s_{pp}}$ = 200 
GeV, was based on previous work on the production of heavy quark states in p-p 
collisions\cite{klm11} which used the color octet model\cite{cl96,bc96,fl96},
which was shown to dominate the color singlet model in studies of $J/\Psi$ 
production at E=200 GeV\cite{nlc03,cln04}. Since the present BNL-RHIC 
experiments cannot measure the $\Upsilon(1S), \Upsilon(2S), \Upsilon(3S)$ 
separately, the $J/\Psi(1S)$ and $\Psi'(2S)$ production was the main study in 
Ref\cite{klm14}.

   An important aspect of the present work is that the $\Psi'(2S)$ and 
$\Upsilon(3S)$ are approximately 50-50 mixtures of standard quarkonium and 
hybrid quarkonium states:
\beq
        |\Psi(2s)>&=& -0.7 |c\bar{c}(2S)>+\sqrt{1-0.5}|c\bar{c}g(2S)>
\nonumber \\
        |\Upsilon(3S)>&=& -0.7 |b\bar{b}(3S)>+\sqrt{1-0.5}|b\bar{b}g(3S)>
 \; ,
\eeq
while the $J/\Psi,\Upsilon(1S),\Upsilon(2S)$ states are 
essentially standard $q \bar{q}$ states\cite{lsk09}. 

  Recently there have been experimental studies with $\sqrt{s_{pp}}$ = 5 TeV of
$\Psi(2S)$ production\cite{aaij16} and $\Upsilon$ production\cite{aaij14}
via p-Pb collisions, and a theoretical estimate of $\Psi,\Upsilon$ 
production\cite{lskdas16} via Pb-Pb collisions.
  In the present work we estimate differential cross sections for $J/\Psi(1S)$,
 $\Psi(2S)$ (charmonium mesons) and $\Upsilon(1S),\Upsilon(2S),\Upsilon(3S)$
(bottomonium mesons) from p-Pb collisions with center of mass energy E= 8 TeV,
which are being planned as future LHCb experiments\cite{2017ECR}. In Sec. II
the expressions for the differential cross sections for $J/\Psi(1S)$ and  
$\Upsilon(1S)$ at 8TeV are derived. In Sec. III the differential cross sections
for $\Psi(2S)$ and $\Upsilon(3S)$, mixed heavy-quark hybrid mesons, as well
as the differential cross sections for the standard heavy-quark mesons are 
given. Due to uncertainty in normalization of the absolute cross sections our 
main prediction is the shapes of the rapidity dependence rather than the 
magnitudes of the cross sections.

\section{$J/\Psi$ and $\Upsilon(1S)$ production via p-Pb collisions 
with $\sqrt{s_{pp}}$ = 8 TeV}

 The differential rapidity cross section for the production of a heavy
quark state with helicity $\lambda=0$ in the color octet model in A-A
collisions is given by

\beq
\label{2}
   \frac{d \sigma_{AA\rightarrow \Phi(\lambda=0)}}{dy} &=& 
   R^E_{AA} N^{AA}_{bin} <\frac{d \sigma_{pp\rightarrow \Phi(\lambda=0)}}{dy}>
\; ,
\eeq
where $R^E_{AA}$ is the product of the nuclear modification factor $R_{AA}$
and $S_{\Phi}$, the dissociation factor after the state $\Phi$ (a charmonium or
bottomonium state) is formed (see Ref\cite{star02}). $N^{AA}_{bin}$ is the number
of binary collisions in the AA collision.
$ <\frac{d \sigma_{pp\rightarrow \Phi(\lambda=0)}}{dy}>$ is the
differential rapidity cross section for $\Phi$ production via nucleon-nucleon
collisions in the nuclear medium:
 
\beq
\label{3}
     < \frac{d \sigma_{pp\rightarrow \Phi(\lambda=0)}}{dy}> &=& 
     A_\Phi \frac{1}{x(y)} f_g(\bar{x}(y),2m)f_g(a/\bar{x}(y),2m) 
\frac{dx}{dy} \; ,
\eeq  
where $a= 4m^2/s$; with $m=1.5$  GeV for charmonium, and 5 GeV for 
bottomonium, and $A_\Phi=\frac{5 \pi^3 \alpha_s^2}{288 m^3 s}<O_8^\Phi(^1S_0)>$
\cite{klm11}. For $\sqrt{s}$ = 8 TeV $a=1.41\times10^{-7}$,  $A_\Phi=4.94\times
10^{-7}$ for charmonium, and  $a=1.56\times10^{-6}$,  $A_\Phi=1.33\times 10^{-8}$
for bottomonium. See Ref\cite{klm11}.

 $\bar{x}$, the effective parton x in a nucleus (A), is\cite{klm14}
\beq
\label{barx}
         \bar{x}(y)&\simeq& x(y) \nonumber \\
   x(y) &=& 0.5 \left[\frac{m}{\sqrt{s_{pp}}}(\exp{y}-\exp{(-y)})+
\sqrt{(\frac{m}{\sqrt{s_{pp}}}(\exp{y}-\exp{(-y)}))^2 +4a}\right] \; ,
\eeq
with $\frac{m}{\sqrt{s_{pp}}}$ = $1.875 \times 10^{-4}$ for charmonium and
$6.25 \times 10^{-4}$ for bottomium

The gluon distribution function $ f_g(x) $ for $\sqrt{s}$ = 8 TeV 
is\cite{klm11}
\beq
\label{fg}
      f_g(x) & \simeq & 275.14 - 6167.6 x + 36871.3 x^2   \; .
\eeq

For the present work we need $R^E_{pPb}$ and $N^{pPb}_{bin}$.
From Ref\cite{sv13} $R^E_{pPb}\simeq 0.8$, while from proton-Nucleus 
experiments\cite{elias78} $N^{pPb}_{bin} \simeq 14.6 $, so $R^E_{AA} N^{AA}_{bin}
\simeq 11.68$
 From these and 
Eqs(\ref{2},\ref{3}) one finds
\beq
\label{dsig-dy}
  \frac{d \sigma_{pPb \rightarrow J/\Psi (\lambda=0)}}{dy} &=& 
    5.78\times 10^{-6} f_g(x(y) f_g(\frac{1.41\times10^{-7}}{x(y)})\times 
\frac{1}{x(y)} \frac{dx(y)}{dy}
\nonumber \\
  \frac{d \sigma_{pPb \rightarrow \Upsilon(1S) (\lambda=0)}}{dy} &=& 
    1.56\times 10^{-7} f_g(x(y) f_g(\frac{1.56\times10^{-6}}{x(y)})\times
\frac{1}{x(y)} \frac{dx(y)}{dy}\; .
\eeq
\newpage
\section{$J/\Psi,  \Psi(2S),  \Upsilon(1S), \Upsilon(2S)$,  
$\Upsilon(3S)$ production via p-Pb \\ collisions with 
$\sqrt{s_{pp}}$ = 8 TeV}

From Ref\cite{klm11}, for the standard model the differential cross secions
$\frac{d \sigma}{dy}$ for $\Psi(2S) = 0.039 \times J/\Psi(1S)$ and 
$\Upsilon(2S) = 0.039 \times  \Upsilon(1S)$, while $\Upsilon(3S) = 0.0064 
\times  \Upsilon(1S)$. Using the mixed hybrid theory\cite{lsk09} Ref\cite{klm11}
found that 
\beq
\label{hyb-norm}
   [\frac{d \sigma_{pp\rightarrow \Psi(2S)}}{dy}]_{mixed\;hybrid} &\simeq& 
2.47 \times [\frac{d \sigma_{pp\rightarrow \Psi(2S)}}{dy}]_{standard} 
\eeq
\beq
  [\frac{d \sigma_{pp\rightarrow \Upsilon(3S)}}{dy}]_{mixed\;hybrid} &\simeq& 
2.47 \times [\frac{d \sigma_{pp\rightarrow \Upsilon(3S)}}{dy}]_{standard}
\nonumber  \; .
\eeq

The differential cross section for $J/\Psi$ is shown in Figure 1. 

\vspace{5cm}

\begin{figure}[ht]
\begin{center}
\epsfig{file=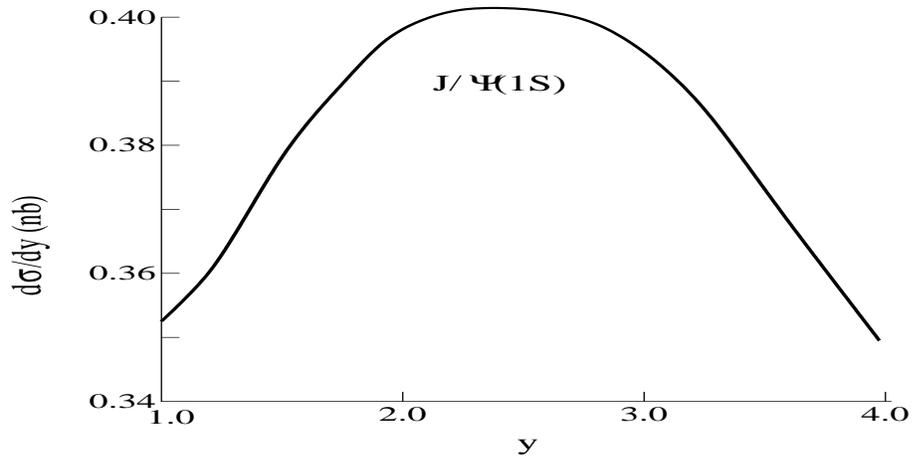,height=6 cm,width=12cm}
\caption{d$\sigma$/dy for  $\sqrt{s_{pp}}$=8 TeV p-Pb collisions 
producing $J/\Psi$ with $\lambda=0$}
\label{Figure 1}
\end{center}
\end{figure}

\newpage
The differential cross sections for $\Psi(2S)$ are shown in Figure 2,
$\Upsilon(1S,2S)$ in Figure 3.
\vspace{5.5cm}

\begin{figure}[ht]
\begin{center}
\epsfig{file=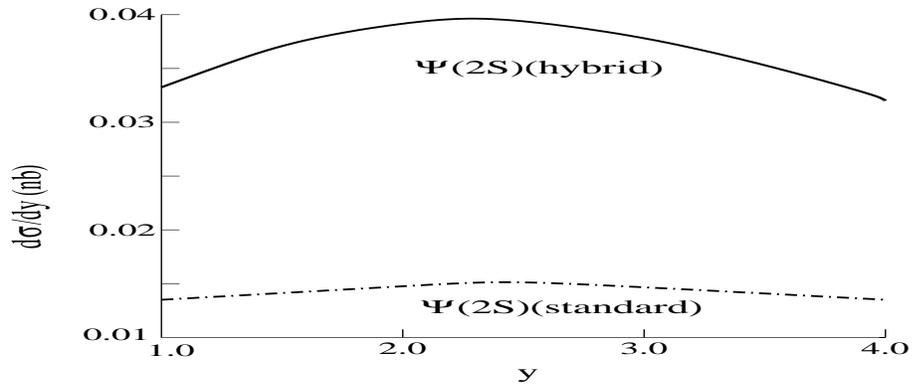,height=5 cm,width=12cm}
\caption{d$\sigma$/dy for$\sqrt{s_{pp}}$=8 TeV p-Pb collisions
producing hybrid, standard $\Psi(2S)$.}
\label{Figure 2}
\end{center}
\end{figure}
\vspace{3.4cm}

\begin{figure}[ht]
\begin{center}
\epsfig{file=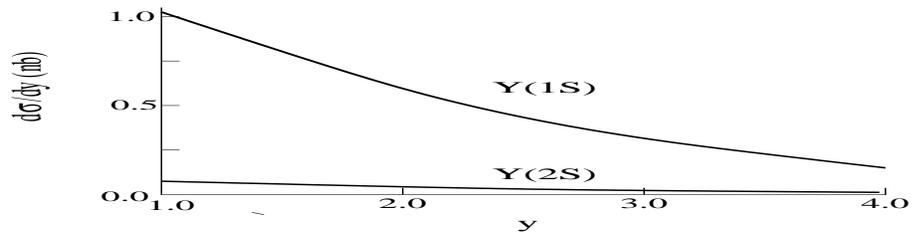,height=3cm,width=12cm}
\caption{d$\sigma$/dy for  $\sqrt{s_{pp}}$=8 TeV p-Pb collisions
producing $\Upsilon(1S)$, $\Upsilon(2S)$.}
\label{Figure 3}
\end{center}
\end{figure}

\newpage
The differential cross sections for $\Upsilon(3S)$ hybrid and standard
(dashed curve) are shown in Figure 4

\vspace{5.5cm}

\begin{figure}[ht]
\begin{center}
\epsfig{file=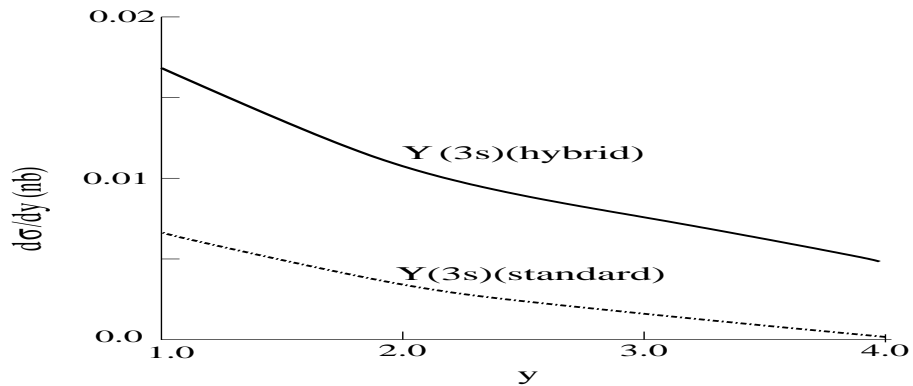,height=5 cm,width=12cm}
\caption{d$\sigma$/dy for$\sqrt{s_{pp}}$=8 TeV p-Pb collisions
producing hybrid, standard $\Upsilon(3S)$.}
\label{Figure 4}
\end{center}
\end{figure}
\section{Conclusions}

In preparing for future LHCb experiments\cite{2017ECR} in which p-Pb collisions 
with center of mass energy E= 8 TeV produce $\Psi$ and $\Upsilon$ states,
we have estimated the differential cross sections for $J/\Psi(1S)$, $\Psi(2S)$,
$\Upsilon(1S)$, $\Upsilon(2S)$, $\Upsilon(3S)$ heavy quark states, with both
the standard model and the mixed hybrid theory for the $\Psi(2S)$ and 
$\Upsilon(3S)$ states. This should help with future experimental studies.
\vspace{1cm}

\Large{{\bf Acknowledgements}}\\
\normalsize
 The author LSK was a visitor at Los Alamos National 
Laboratory, Group P25 while this research was being carried out. The author
thanks Dr. Cesar L. Da Silva for  suggesting this research and helpful
discussions.

\end{document}